\documentclass[twocolumn,preprintnumbers,amsmath,amssymb]{revtex4}
\usepackage{graphicx}% Include figure files
\usepackage{dcolumn}% Align table columns on decimal point
\usepackage{bm}% bold math
\usepackage{epsfig}

\usepackage{dcolumn}% Align table columns on decimal point
\usepackage{bm}% bold math
\usepackage{color,graphicx}
\usepackage{epsfig}
\usepackage{xcolor}
\usepackage{hyperref}

\newcommand{\gevsq}{~(\mathrm{GeV}/{\it c})^2}
\newcommand{\eq}[1]{Eq.(\ref{#1})}

\begin{document}

\title{The EMC Effect of Tritium and Helium-3
 from the JLab MARATHON Experiment}

\author
{ 
 {D.~Abrams},$^{1}$ 
 {H.~Albataineh},$^{2}$
 {B.~S.~Aljawrneh},$^{3}$
 {S.~Alsalmi},$^{4,5}$
 {D.~Androic},$^{6}$
 {K.~Aniol},$^{7}$
 {W.~Armstrong},$^{8}$
 {J.~Arrington},$^{8,9}$
 {H.~Atac},$^{10}$
 {T.~Averett
% ...
} $^{27}$
}
\vskip1.145cm
\affiliation{$^{1}$University of Virginia, Charlottesville, Virginia 22904, USA}
\affiliation{$^{2}$Texas A \& M University, Kingsville, Texas 78363, USA}
\affiliation{$^{3}$North Carolina A \& T State University, Greensboro, North Carolina 27411, USA}
\affiliation{$^{4}$Kent State University, Kent, Ohio 44240, USA}
\affiliation{$^{5}$King Saud University, Riyadh 11451, Kingdom of Saudi Arabia}
\affiliation{$^{6}$University of Zagreb, 10000 Zagreb, Croatia}
\affiliation{$^{7}$California State University, Los Angeles, California 90032, USA}
\affiliation{$^{8}$Physics Division, Argonne National Laboratory, Lemont, Illinois 60439, USA}
\affiliation{$^{9}$Lawrence Berkeley National Laboratory, Berkeley, California 94720, USA}
\affiliation{$^{10}$Temple University, Philadelphia, Pennsylvania 19122, USA}
% ...
\affiliation{$^{27}$Stony Brook, State University of New York, New York 11794, USA}
% ...

\date{\today}% It is always \today, today,

\begin{abstract}
\hspace*{1.0in} The Jefferson Lab Hall A Tritium Collaboration \\

%\hspace*{1.5in} To be submitted to {\it Physical Review Letters} \\

Measurements of the EMC effect in the tritium and helium-3 mirror nuclei are reported.  The data
were obtained by the MARATHON Jefferson Lab experiment, which performed deep inelastic electron
scattering from deuterium and the three-body nuclei, using a cryogenic gas target system and the
High Resolution Spectrometers of the Hall A Facility of the Lab.  The data cover the Bjorken $x$
range from 0.20 to 0.83, corresponding to a squared four-momentum transfer $Q^2$ range from 2.7 to
$11.9\gevsq$, and to an invariant mass $W$ of the final hadronic state greater than 1.84 GeV/${\it c}^2$.
The tritium EMC effect measurement is the first of its kind.  The MARATHON experimental results are
compared to results from previous measurements by DESY-HERMES and JLab-Hall C experiments, as well
as with few-body theoretical predictions. 

\end{abstract}

\maketitle

The European Muon Collaboration (EMC) discovered a significant suppression of the electron deep
inelastic scattering (DIS) cross section (or equivalently the structure function $F_2$) for iron per
nucleon with respect to that of deuterium for the Bjorken scaling variable $x$ from 0.3 to 0.7, 
corresponding to the quark-valence region~\cite{au83}.  Bjorken $x$ is defined in the lab frame as
$x=Q^2/[2M(E-E^\prime)]$, where $M$ is the nucleon mass, and $E$ and $E^\prime$ are the incident 
and scattered lepton energies in the scattering from the nucleus.  This effect, named \emph{EMC effect}, 
was confirmed by a reanalysis of older SLAC data~\cite{bo83}, and by various experiments with electron
and muon beams~\cite{go94,am95,ai03,se09}.  Nuclear effects are commonly studied using the ratio
$R_{A}=\sigma_A/\sigma_d$ of cross sections for scattering off a nucleus $A$ and deuterium $d$,
normalized per nucleon. In the valence quark region $0.3<x<0.6$, $R_A$ is approximately a
linear function of $x$ with negligible $Q^2$ dependence.  The slope $\mathrm d R_A/\mathrm d x$
in this region depends on the nuclear target and its value increases with the nuclear mass
number $A$.  Nuclear effects on $R_A$ have also been studied in other kinematical regions.  (For
a review of data and models see Refs.~\cite{ar94,ge95,no03,ma14,he17}.)

In the infinite momentum frame, $x$ can be interpreted in DIS as the fraction of the target nucleon's
momentum carried by the struck quark.  Momentum conservation suggests that if the valence quark
fraction distribution is suppressed in nuclei then the corresponding lower-$x$ fraction in the total
distribution should be enhanced.  Quite a few models have been suggested to explain the redistribution
of missing valence light-cone momentum in nuclei between bound nucleons and non-nucleon degrees of
freedom in nuclei such as nuclear pions, nucleon resonances, multi-quark clusters, and change of the
quark-gluon confinement scale in nuclear environment.  (For a review see 
Refs.~\cite{ar94,ge95,no03,ma14,he17}.)  It has been known since the 1970s that
smearing the nucleon structure function with the nuclear momentum distribution (due to Fermi motion)
results in an enhancement of the nuclear structure functions at high $x$~\cite{we72,bo80}.  The
available EMC-effect data have shown this enhancement, but also a significant suppression for $x$ up
to about 0.7.  A revision of the Fermi motion correction to include the effect of the nuclear
binding allows a reduction of the discrepancy between calculations and data~\cite{ak85}.
Further refinements and quantitative studies of the Fermi motion and nuclear binding with 
a realistic nuclear spectral function including a high-momentum component, can explain about half
of the observed EMC effect at its maximum value around $x\sim0.7$~\cite{ku89,ci89,ku06}, somewhat
underestimating the value of the slope $\mathrm d R_A/\mathrm d x$ for $0.3<x<0.7$.

Since bound nucleons are off-mass-shell due to the nuclear binding, their invariant mass squared
is, for kinematic reasons, less than $M^2$.  This off-shell effect results in a nuclear modification 
to the structure of the bound nucleons, after averaging with the nuclear energy-momentum 
distribution~\cite{ku94}. In the theoretical model of Ref.~\cite{ku06} this effect is addressed in terms
of a dimensionless function $\delta f(x)$ describing the relative off-shell correction to the nucleon 
$F_2$ structure function.  There it was shown that the EMC effect can be described with high accuracy
over the complete kinematic region covered by existing data using the same $\delta f(x)$ function for 
bound protons and neutrons. Predictions based on this assumption were verified with a broad range of
data from a variety of high-energy processes~\cite{ku10,ku14,ru16}.  Further study of a possible
isospin dependence~\cite{ku14,tr19} of this correction requires the use of nuclei with a large neutron
or proton excess like $^3$H and $^3$He.  Nuclear modifications of various types of the bound nucleon
structure in the valence quark region are also present in a number of different models~\cite{fr85,cl06,se20}. 
Other nuclear effects, such as corrections from meson-exchange currents and the propagation of the
hadronic (quark-gluon) component of the virtual intermediate photon in the nuclear environment are
relevant in the small $x$ region~\cite{ar94,ge95,no03,ku17}.

A crucial step in understanding the origin of the EMC effect is a comparison of realistic calculations
of the structure functions of the lightest nuclei, deuterium [$^2\text{H}$], helium-3 [$^3\text{He}$~$(h)$], 
and tritium [$^3\text{H}$~$(t)$], with precision measurements.  In this Letter we report the measurement
of the EMC effect of the $A=3$ mirror nuclei by the MARATHON Jefferson Lab (JLab) experiment~\cite{pe10}, 
which previously determined the ratio of the proton~($p$) and neutron~($n$) $F_2$ structure functions, 
$F_2^n/F_2^p$, from DIS measurements off tritons ($t$) and helions ($h$)~\cite{ab22}.  MARATHON used the
Continuous Electron Beam Accelerator and the Hall A Facility~\cite{al04} of JLab.

Electrons scattered from $d$, $h$, and $t$ nuclei in high-pressure, cryogenic gas target cells~\cite{ho10},
cooled to a temperature of 40~K, were detected in the Left and Right High Resolution Spectrometers (HRS) 
of the Hall~\cite{al04}.  The incident-beam energy was fixed at 10.59 GeV, and the beam current ranged
from 14.6 to 22.5~$\mu$A. The Left HRS was operated at a fixed momentum of 3.1 GeV/{\it c}, placed at angles
between $16.8^{\circ}$ and $33.6^{\circ}$.  The Right HRS was operated at a single setting of 2.9 GeV/{\it c} 
and $36.1^{\circ}$.  In each HRS system, particles were detected using two planes of scintillators for event
triggering, a pair of drift chambers for track reconstruction, and a gas threshold Cherenkov counter
and a lead-glass calorimeter for electron identification.  The target cells were cycled many times in the 
beam for each kinematic setting in order to minimize effects of possible drifts of the beam diagnostic or 
other instrumentation ({\it e.g.} the beam current monitors).  Essential information for the experimental 
apparatus has been provided in Ref.~\cite{ab22}.  Additional detailed information on the Hall A spectrometer 
facility, and the associated beam instrumentation with calibrations, as used in MARATHON, are given in 
Refs.~\cite{ba19,ha20,ku19,li20,ny20,su20}.

All events identified as electrons originating from the gas inside each target cell were binned by Bjorken
$x$, resulting in the formation of an electron yield, equal to the number of scattered electrons for each
bin divided by the number of incident beam electrons and of gas target nuclei per unit area, as described
in Ref.~\cite{ab22}.  The ratio of the yields for two targets is equivalent to the ratio of their cross
sections, because the associated identical effective target lengths~\cite{ab22} and solid angles cancel
out in the latter ratio.  The overall electron detection efficiency, close to unity ($\sim 0.985$), was
found to be independent of the target cell at all kinematics, so it also cancels out in the ratios of the 
yields.  Several multiplicative correction factors were applied to the individual target yields.  The
correction for
i) computer dead-time ranged from 1.001 to 1.065,
ii) target density change due to beam heating effects from 1.066 to 1.112,
iii) falsely-reconstructed events originating from the end-caps from 0.973 to 0.998,
iv) events originating from pair symmetric processes from 0.986 to 0.999,
v) radiative effects from 0.853 to 1.167,
vi) beta decay of tritons to helions (applicable only to the tritium yield) from 0.997 at the 
beginning to 0.989 at the end of the experiment,
vii) Coulomb distortion effects  from 0.997 to 1.000, and
viii) bin-centering adjustment  from 0.995 to 1.001.
In the above, the ranges refer to the $^3$He, $^3$H, and $^2$H gas yields.  A cross section model from
Refs.~\cite{ku06,ku10} was adopted for the Coulomb correction (which used the $Q^2$-effective approximation 
as outlined in Ref.~\cite{ub71}), and for the bin-centering correction.

The corrections to the $h/d$ and  $t/d$ cross section ratios from each effect listed above become minimal, 
and in some cases, so do the associated systematic uncertainties.  For example, the radiative effect 
correction ranges from 1.000 to 1.004 and 1.006 to 1.012, respectively.  The dominant point-to-point 
systematic uncertainties for the yield ratios are those from the beam-heating gas target density changes 
[$\pm(0.1\%$-$0.5\%)$], the radiative correction [$\pm(0.25\%$-$0.45\%)$], and the choice of spectrometer 
acceptance limits ($\pm0.2\%$).  The total point-to-point uncertainty ranged from $\pm0.46\%$ to $\pm0.49\%$ 
for the $h/d$ cross section ratio, and $\pm0.34\%$ to $\pm0.47\%$ for the $t/d$ ratio.  Details on the
determination of the yields, the associated corrections and uncertainties, and other relevant subjects
can be found in Refs.~\cite{ba19,ha20,ku19,li20,ny20,su20}.

The experiment also collected DIS data for the proton, in the $x$ range from 0.20 to 0.38, for normalization 
purposes.  The resulting $\sigma_d/\sigma_p$ ratio measured by MARATHON is in excellent agreement with the 
{\it reference} measurements of the seminal SLAC-E49b and E87 experiments~\cite{bo79}, as shown in  
Ref.~\cite{ab22}.  The $\sigma_d/\sigma_p$ data from MARATHON allowed for an accurate determination of the 
$R_{np}=\sigma_n/\sigma_p$ ratio from the relation~\cite{ku10,pa01} $R_{np}=(\sigma_d/\sigma_p)/R_d-1$, 
where $R_d=\sigma_{d}/(\sigma_p+\sigma_n)$.

In the extraction of $R_{np}$ from the MARATHON $^3$He and $^3$H data, it was realized~\cite{ab22}
that the $\sigma_h/\sigma_t$ ratio had to be normalized by a factor of 1.025, a result of requiring
the equality of the $R_{np}$ values extracted from $\sigma_h/\sigma_t$ and $\sigma_d/\sigma_p$ in
the vicinity of $x=0.3$.  In this work we follow the same approach by requiring that the $R_{np}$
value extracted individually from $\sigma_t/\sigma_d$ and $\sigma_h/\sigma_d$ be equal to that
extracted from $\sigma_d/\sigma_p$ in the vicinity of $x=0.3$, where nuclear corrections are minimal.
We define the EMC-type ratios for the cross sections of $^3$He~($h$) and $^3$H~($t$) as
$R_h = \sigma_h/(2 \sigma_p + \sigma_n)$ and $R_t = \sigma_t/(\sigma_p + 2 \sigma_n)$, respectively.
Then, the double ratios ${\cal R}_{hd} = R_h/R_d$ and ${\cal R}_{td} = R_t/R_d$ allow for the determination
of $R_{np}$ in two separate ways:
\begin{align}
\label{super}
R_{np} = \frac{2\mathcal R_{hd}(\sigma_d/\sigma_h) - 1}{1-\mathcal R_{hd}(\sigma_d/\sigma_h)}=
%\label{super:td}
%R_{np} &= 
\frac{\mathcal R_{td}(\sigma_d/\sigma_t) - 1}{1-2\mathcal R_{td}(\sigma_d/\sigma_t)},
\end{align}
once the ratios $\sigma_h/\sigma_d$ and $\sigma_t/\sigma_d$ have been measured experimentally, and the 
ratios ${\cal R}_{hd}$ and ${\cal R}_{td}$ have been theoretically calculated with a reliable model.

Predictions for the $R_d$, ${\cal R}_{hd}$, and ${\cal R}_{td}$ ratios were obtained prior to the
analysis of the MARATHON data from the theoretical model of Kulagin and Petti~(K-P)~\cite{ku10,ku06},
which provides a very good description of the EMC effect for all known targets (for a review see 
Ref.~\cite{ku17}). This model includes a number of nuclear effects out of which the major correction
for the relevant kinematics comes from the smearing effect with the nuclear energy-momentum distribution, 
described in terms of the nuclear spectral function, together with an off-shell correction to the bound
nucleon cross sections~\cite{ku06}.  The underlying nucleon structure functions come from a
global QCD analysis~\cite{al07}, which was performed up to NNLO approximation in the strong
coupling constant including target mass corrections~\cite{ge76} as well as those due to higher-twist 
effects~(OPE~\cite{ge76}).  For the spectral functions of the ${^3}$H and ${^3}$He nuclei, the results of
Ref.~\cite{pa01} have been used, while for the $^2$H wave function of the Argonne AV18 nucleon-nucleon
interaction~\cite{av18} was applied.  In order to evaluate theoretical uncertainties, the ${^3}$He
spectral function of Ref.~\cite{sc92} and the Bonn $^2$H wave function of Ref.~\cite{bonn} was used.
Reasonable variations of the high-momentum part of the nucleon momentum distribution in ${^3}$H
and ${^3}$He were considered, and uncertainties in the off-shell correction of Ref.~\cite{ku06},
as well as in the nucleon structure functions of Ref.~\cite{al07}, were accounted for~\cite{kppc}.

The comparison of $R_{np}$ as extracted from the measured $\sigma_h/\sigma_d$, $\sigma_t/\sigma_d$, and
$\sigma_d/\sigma_p$ ratios was actually done at $x=0.31$, where nuclear corrections are not expected
to contribute to isoscalar nuclear ratios like $\mathcal R_{hd}$, $\mathcal R_{td}$, and $R_d$.
This expectation is based on the experimental data for $A\geq 3$ nuclei~\cite{ai03,se09,go94,am95}
in the range $0.25 \leq x \leq 0.35$, taking into account the quoted normalization uncertainties therein. 
This approach is also in line with the results of Refs.~\cite{ku10,we11}.  The K-P model predicts a value
of 1.000, 1.000, and 1.000 at $x=0.31$ for ${\cal R}_{hd}$, ${\cal R}_{td}$, and $R_d$, with uncertainties
of $\pm0.38\%$, $\pm0.42\%$ and $\pm0.20\%$, respectively.  The above value for $R_d$ is in very good
agreement with the independent analyses of Refs.~\cite{go94,gr15,al17,se20}.  The values of 
$\sigma_{d}/\sigma_p$, $\sigma_h/\sigma_d$, and $\sigma_t/\sigma_d$ at $x=0.31$ were determined by weighted
fits to the three corresponding sets [$x$ between 0.20 and 0.83~(0.20 and 0.38) for 
$\sigma_{h}/\sigma_d$ and $\sigma_t/\sigma_d$~($\sigma_{d}/\sigma_p$)], which included statistical 
and point-to-point uncertainties added in quadrature.

\begin{figure} [htb]
\centering
\includegraphics[width=90mm, angle=0]{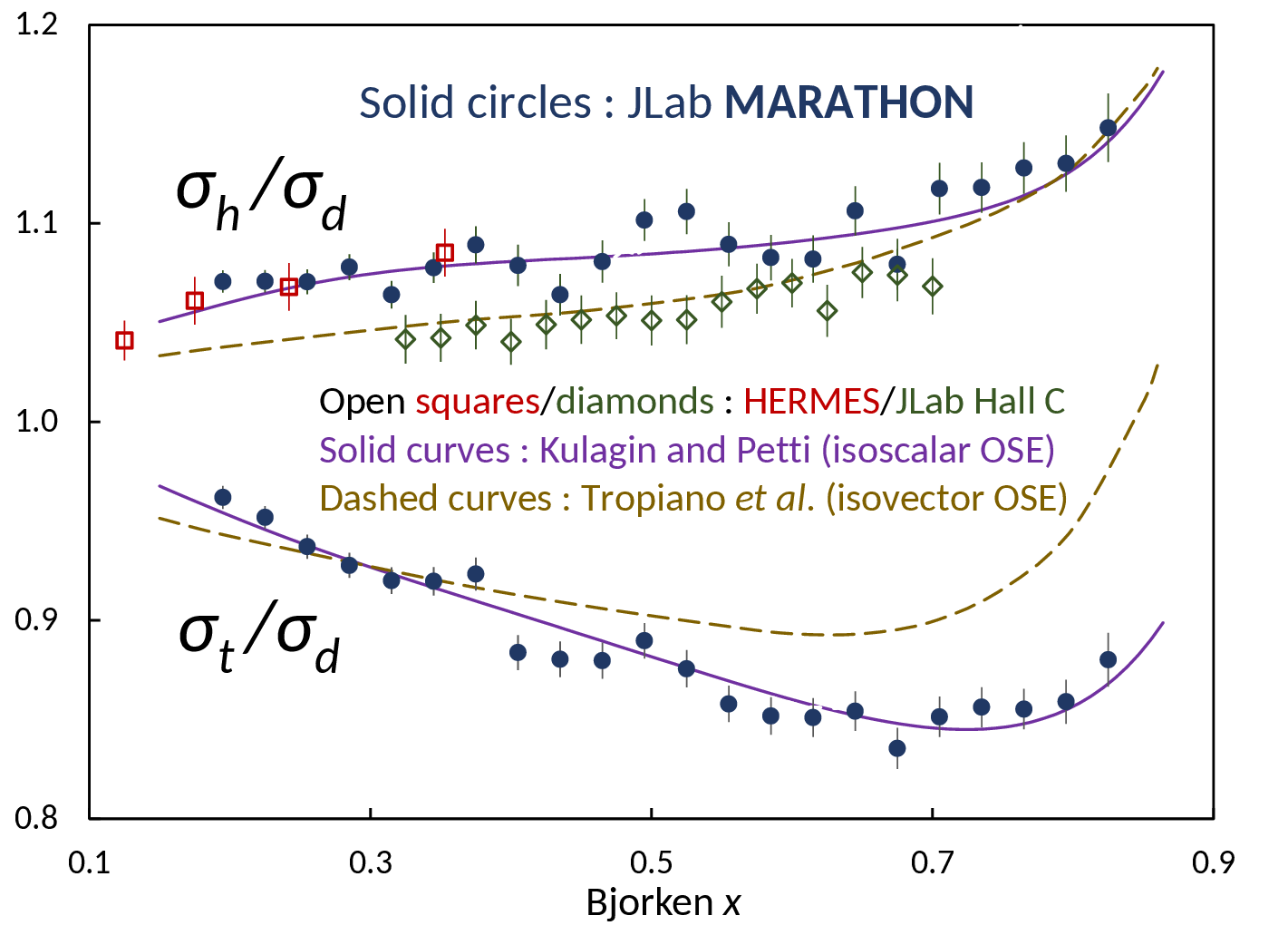}
\caption{\label{fig1}
The MARATHON results on the $^3\text{He}/{}^2\text{H}$ and  $^3\text{H}/{}^2\text{H}$ DIS cross
section ratios {\it versus} Bjorken $x$.  The solid curves are the prediction of the K-P model
(with isoscalar OSE)~\cite{ku10,kppc}, and the dashed curves are the result of Tropiano {\it et al.}
(with isovector OSE)~\cite{tr19}.  Also shown are data from the DESY-HERMES experiment~\cite{ai03},
and final results from JLab Experiment E03-103~\cite{se09} (see text).  The error bars include
statistical and point-to-point systematics uncertainties (added in quadrature).}
%\vspace* {-.20in}
\end{figure}

In order to match the $\sigma_n/\sigma_p$ values found using the three different sets of nuclei, the
$\sigma_h/\sigma_d$ and $\sigma_t/\sigma_d$ ratios at $x=0.31$ had to be normalized by a multiplicative
factor of 1.021$\pm$0.005 and 0.996$\pm0.005$, respectively.  These two factors are perfectly consistent
with the normalization factor of 1.025$\pm$0.007 of the $\sigma_h/\sigma_t$ ratio, as determined similarly
in Ref.~\cite{ab22}.  All values for the $\sigma_h/\sigma_d$ and $\sigma_t/\sigma_d$ ratios reported and
further used in this work have been normalized using these two factors.  The normalized ratios' values 
are given in Tables 1 and 2 of the online supplemental File, together with associated uncertainties, and 
plotted in Fig.~\ref{fig1}.  As a matter of convention, which will be followed for the remainder of this 
work, the ratios have been adjusted so that the cross sections are per nucleon.  

The data are compared to the theoretical predictions of the K-P model~\cite{ku10,kppc} and the model of
Tropiano {\it et al.}~(TEMS)~\cite{tr19}.  Both K-P and TEMS predictions are based on a nuclear convolution
approach~\cite{ku06,ku08,ku10}, but they involve different assumptions.  As mentioned above, K-P use the
proton and neutron structure functions from a global QCD fit~\cite{al07} and the relative off-shell
effect~(OSE) correction from Ref.~\cite{ku06}.  The TEMS group employs the results of the CJ15 
analysis~\cite{cj15} with the OSE correction adjusted from a fit to JLab Hall C $\sigma_h/\sigma_d$ 
data~\cite{se09}, allowing for different off-shell modifications for bound protons and neutrons.  The 
latter data are also shown in Fig. 1 for $W^2$ values greater than 3.4~$({\rm GeV}/{\it c}^2)^2$.  Here, 
it should be noted that in order for the Hall C data to provide values of $\sigma_n/\sigma_p$ which would 
match those of MARATHON, they should be adjusted upwards by about 2.5$\%$.  Such a requirement would make 
the two $\sigma_h/\sigma_d$ data sets mutually perfectly consistent.  The MARATHON data are in excellent 
agreement with the K-P prediction over the entire measured range of $x$, as quantified by a $\chi^2$ per 
degree of freedom of 1.0, and with overlapping data from the HERMES experiment~\cite{ai03}, also shown 
in Fig. 1.

\begin{figure} [htb]
\begin{center}
\includegraphics[width=88mm, angle=0]{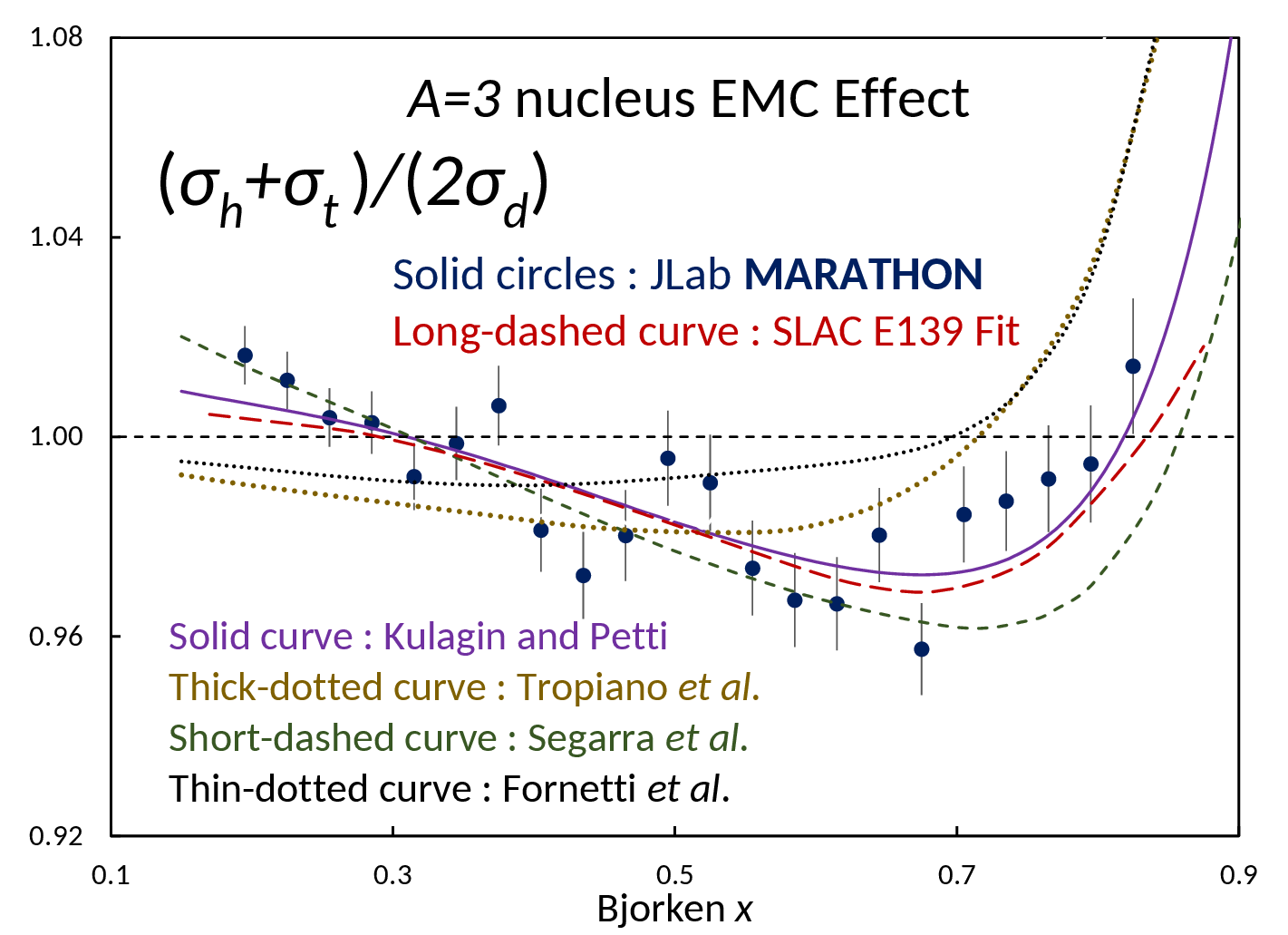}
\end{center}
\caption{\label{fig2}
The $A=3$ EMC effect in a isoscalar combination of the ${}^3\text{He}$ and ${}^3\text{H}$ cross sections
{\it versus} Bjorken $x$. The error bars include statistical and point-to-point systematics uncertainties. 
The solid curve is the prediction of the K-P model~\cite{ku10,kppc} (with isoscalar OSE), the thick-dotted,
dashed, and thin-dotted curves are results from Ref.~\cite{tr19} (with isovector OSE), Ref.~\cite{se20}, and
Ref.~\cite{fo24}, respectively.  The long-dashed curve shows the $A$-dependent SLAC-E139
parameterization~\cite{go94}.}
%\vspace* {-.20in}
\end{figure}

The arithmetic mean of $\sigma_h/\sigma_d$ and $\sigma_t/\sigma_d$ provides a model-independent
determination of the average isoscalar EMC effects of three-body mirror nuclei $^3$He and $^3$H. 
The resulting isoscalar ratio $(\sigma_h+\sigma_t)/(2\sigma_d)$ values and associated uncertainties
are listed in Table 3 of the online supplemental file, and plotted in Fig.~\ref{fig2} along with statistical
and point-to-point systematic uncertainties added in quadrature.  Also shown are the result of the
SLAC-E139 parametrization of the EMC effect, for $A=3$, in terms of $\ln(A)$~\cite{go94}, and the predictions
of K-P~\cite{ku10,kppc} and TEMS~\cite{tr19}, as well as the  results of calculations by Segarra
{\it et al.}~\cite{se20} and Fornetti {\it et al.}~\cite{fo24}.  The former is based on a parameterization of
the EMC effect in terms of the fraction of the nuclear high-momentum component generated by short-range 
correlations.  The latter is based on a Poincaré covariant approach, using light-front Hamiltonian dynamics 
and wave functions obtained from modern nuclear interactions.

To obtain the isoscalar EMC effect separately for the $^3$H and $^3$He nuclei, the ratios $\sigma_t/\sigma_d$
and $\sigma_h/\sigma_d$ must be corrected for the neutron and proton excess, respectively.  This is achieved
by multiplying them by the customary factor
\begin{equation}\label{fiso}
F_\text{iso} = \frac{A(1+R_{np})}{2[Z+(A-Z)R_{np}]},
\end{equation}
where $Z$ is the nuclear atomic number.  The $R_{np}=\sigma_n/\sigma_p$ values used in \eq{fiso} are the ones
measured by the MARATHON experiment and reported in Ref.~\cite{ab22}.  The values of the correction can be
inferred from the information provided in Tables 1, 2, 4 and 5 of the online supplemental File.  For $^3$He, 
they range from 0.949 (lowest $x$) to 0.888 (highest $x$).  For $^3$H, they range from 1.057 (lowest $x$) to 
1.144 (highest $x$).  Using the MARATHON-extracted $R_{np}$ ({\it i.e.} substituting $R_{np}$=$F_2^n/F_2^p$ as 
given by Eq.~(2) of Ref.~\cite{ab22} in the above Eq.~(2)) allows us to cast the two isoscalar EMC ratios as 
follows:
\begin{align}
\left( \sigma_h / \sigma_d \right)_\text{iso} &=
\tfrac12\left[\sigma_h/\sigma_d + {\cal R}_{ht}(\sigma_t /\sigma_d)\right], 
%\quad
\\
\left( \sigma_t / \sigma_d \right)_\text{iso} &=
\tfrac12\left[\sigma_t / \sigma_d + (\sigma_h/ \sigma_d)/\mathcal R_{ht}\right],
\end{align}
where the ``super-ratio" $\mathcal R_{ht}=R_h/R_t$ is the ratio of the previously defined $R_h$ and $R_t$
ratio quantities, and for which the K-P model prediction is used~\cite{ku10,ab22}.
The values of $\mathcal R_{ht}$ are listed, along with their estimated theory uncertainties, in Ref.~\cite{ab22}, 
where it can be seen that the deviation of $\mathcal R_{ht}$ from unity is well below 1\% for most points, 
with a maximal value reaching 1.25\%.  Note that $({\sigma_h/\sigma_t})_\text{iso} = {\cal R}_{ht}$, and in 
the limit $\mathcal R_{ht}=1$ we have identical individual isoscalar EMC-effects for $^3$H and $^3$He.

\begin{figure} [htb]
\centering
\includegraphics[width=88mm, angle=0]{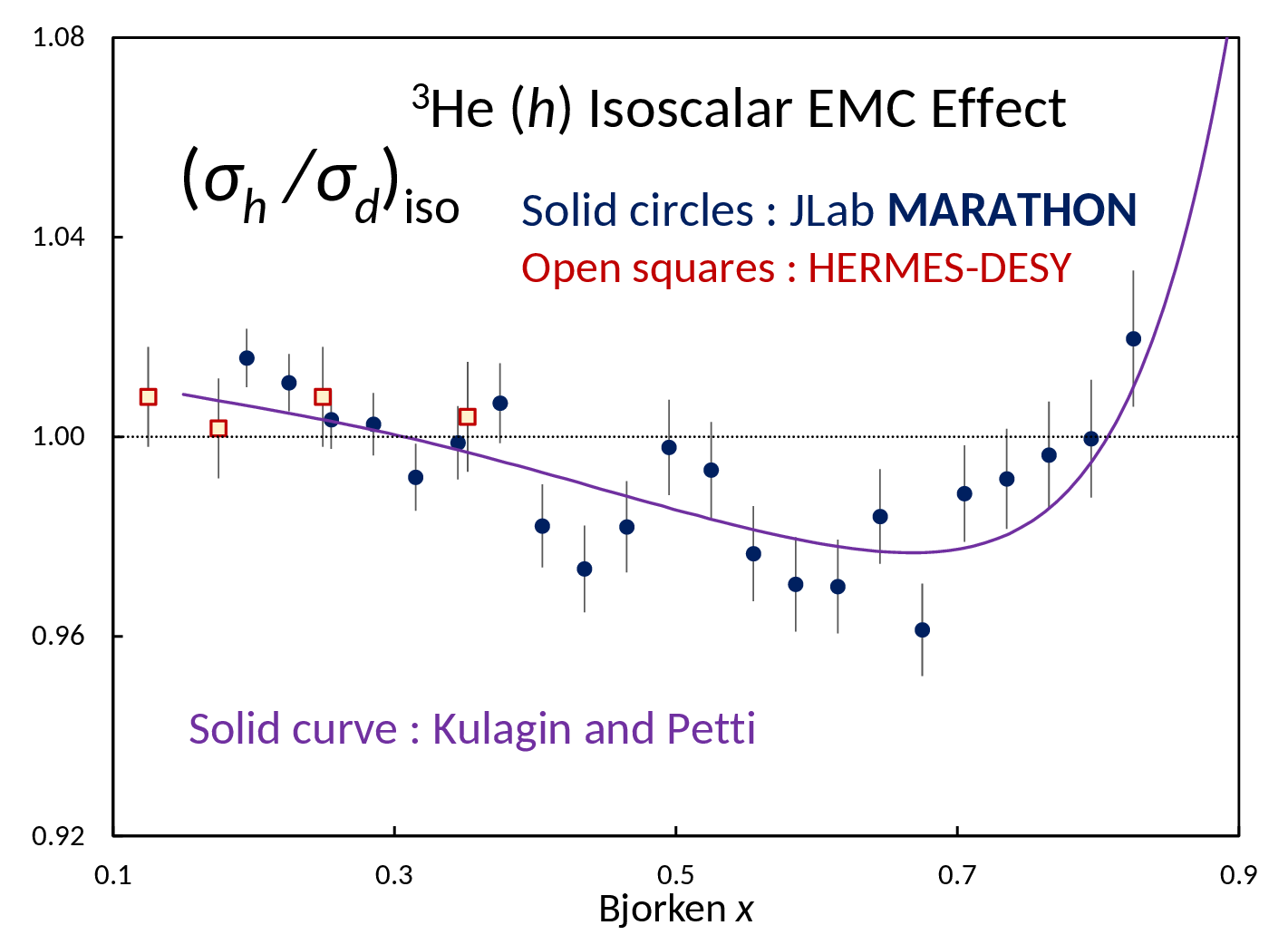}
\caption{\label{fig3}
The $^3\text{He}/{}^2\text{H}$ cross section ratio from the MARATHON
experiment corrected for isoscalarity {\it versus} Bjorken $x$.
Also shown are the results from the DESY-HERMES experiment~\cite{ai03,ga03}. 
The error bars include statistical and point-to-point systematic uncertainties.
The solid curve is the prediction of the K-P model~\cite{ku10,kppc}.
}
%\vspace* {-.20in}
\end{figure}

\begin{figure} [htb]
\centering
\includegraphics[width=88mm, angle=0]{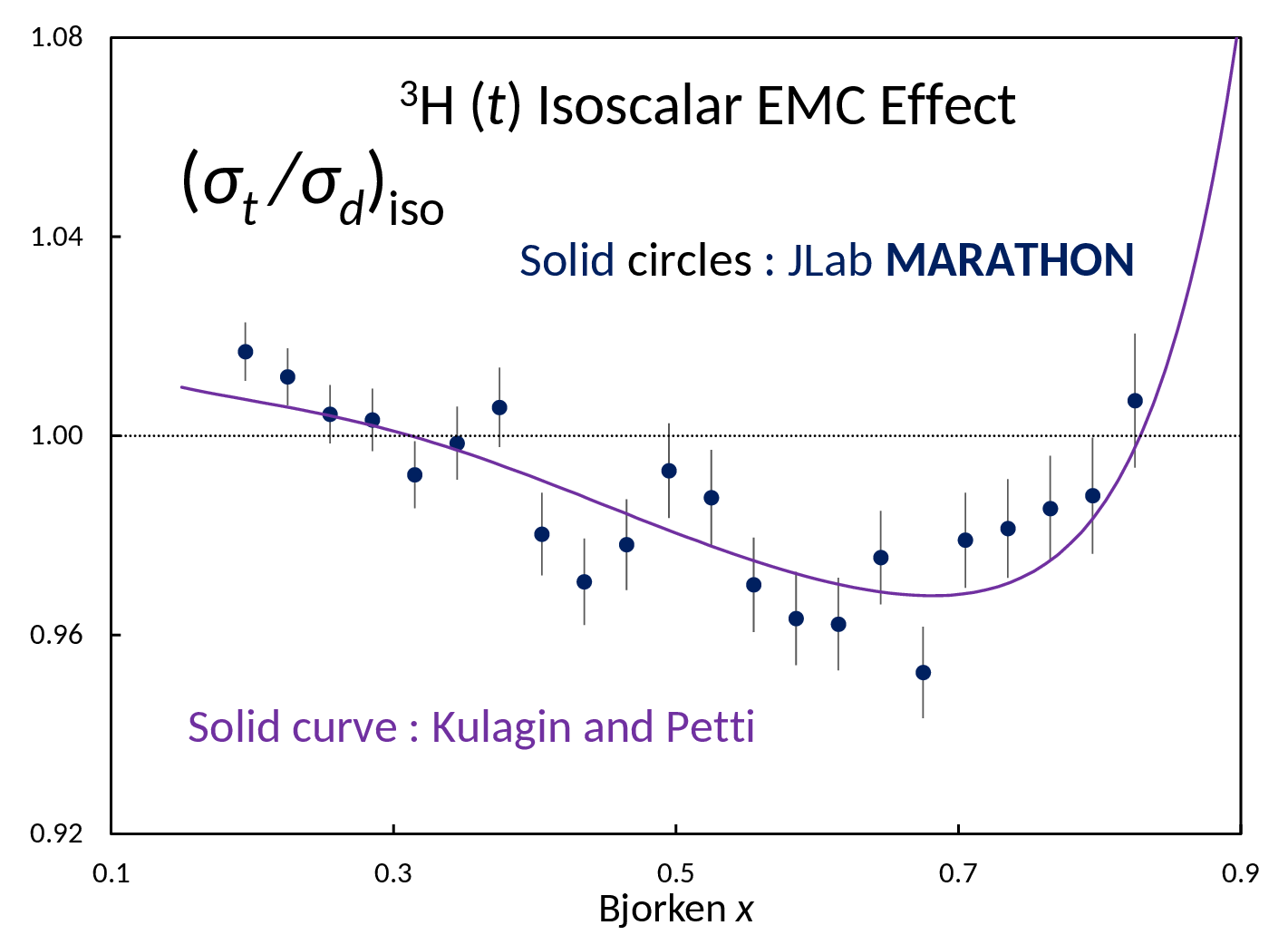}
\caption{ \label{fig4} {The $^3\text{H}/{}^2\text{H}$ cross section ratio from the MARATHON experiment
corrected for isoscalarity {\it versus} Bjorken $x$.  The error bars include statistical and point-to-point
systematics uncertainties. The solid curve is the prediction of the K-P model~\cite{ku10,kppc}.} }
%\vspace* {-.20in}
\end{figure}

The measured values of $(\sigma_h/\sigma_d)_{\rm iso}$ and $(\sigma_t/\sigma_d)_{\rm iso}$ of the individual
EMC effects of the two $A=3$ nuclei are given in Tables 4 and 5 of the online supplemental file, together with
associated uncertainties, and plotted in Figs.~\ref{fig3} and \ref{fig4}, respectively.  Since the ratios
${\sigma_h/\sigma_d}$ and ${\sigma_t/\sigma_d}$ are correlated, the uncertainties of the three $A=3$, $^3$H, 
and $^3$He EMC effects have been determined by a Monte Carlo simulation, where it has been estimated that one
half of both the point-to-point and overall scale uncertainties of the two ratios are correlated.  A model is
used for each EMC effect and then its uncertainty (statistical or systematic) is the root mean square deviation
between the model and the randomized simulated values.  The essence of such simulation is that the same random
deviation is used to account for the uncertainty of the deuteron yield, which appears twice in the functional
form of each ratio.  Also shown in the Figs. are the predictions of K-P model~\cite{ku10,kppc}.  Fig.~\ref{fig3} 
also shows data from the HERMES experiment~\cite{ai03}, as listed in Ref.~\cite{ga03}, which are in excellent 
agreement with the MARATHON data.  The JLab Hall C data~\cite{se09}, which are consistent with the MARATHON 
data within overall normalizations, are not shown as they have been determined with different isoscalarity 
correction.  (The isoscalarity corrections of the MARATHON and HERMES data are mutually consistent.)  

It is customary to extract the slope of $(\sigma_A/\sigma_d)_\text{iso}$ in the $x$ range between 0.3 and 0.7 
from EMC measurenets, assuming that the effect follows there an approximate linear behavior.  A linear fit 
to the MARATHON data including statistical and point-to-point systematic uncertainties results in the values 
of $-0.085\pm0.037$ and $-0.10\pm0.04$ for $^3$He and $^3$H, respectively.  The $^3$He slope value is similar 
to the $-0.085\pm0.027$ one from JLab E03-013 experiment~\cite{se09}, although it uses different isoscalarity 
correction factor values than MARATHON.

In summary, the MARATHON experiment has provided a precise measurement of the EMC effect for the three-body 
nuclei $^3$He and  $^3$H individually, as well as for their $A=3$ isoscalar combination, at large four 
momentum transfers in the DIS regime, with $0.20<x<0.83$.  The extracted EMC effect for an $A=3$ nucleus are 
consistent with the $A$-dependent SLAC-E139 parametrization based on measurements for $A\ge4$ nuclei~\cite{go94}.
The new MARATHON data are in agreement with theoretical predictions in which nuclear corrections originate 
from  the energy-momentum distribution of bound nucleons together with an off-shell modification of their 
internal structure~\cite{ku06,ku10}, but they do not provide evidence for a sizable isovector EMC effect 
component as argued in Ref.~\cite{tr19}.  The new data are expected, in general, to provide unique input for 
the study of the partonic structure of the few-body nuclear systems. 

We acknowledge the outstanding support of the staff of the Accelerator Division and Hall A Facility of
JLab, and work of the staff of the Savannah River Tritium Enterprises and the JLab Target Group.
We thank Drs. M.~E.~Christy and Y.~Kolomensky for useful discussions on the optical properties of the HRS 
systems, and the interpretation of the data, respectively.  We are grateful to Dr. W.~Melnitchouk for his 
contributions to the development of the MARATHON proposal, and to Dr. A.~W.~Thomas for many valuable 
discussions on and support of the MARATHON project since its inception.
This material is based upon work supported by the U.S. Department of Energy (DOE), Office of Science,
Office of Nuclear Physics under contracts DE-AC05-06OR23177 and DE-AC02-06CH11357.
This work was also supported by DOE contract DE-AC02-05CH11231, DOE award DE-SC0016577, DOE award DE-SC0010073,
National Science Foundation awards NSF-PHY-1405814 and NSF-PHY-1714809, the Kent State University Research
Council, the Pazy Foundation and the Israeli Science Foundation under grants 136/12 and 1334/16, Grant 
21AG-1C085 by the Science Committee of the Republic of Armenia, and the Italian Institute of Nuclear Physics.

Notice: Authored by Jefferson Science Associates, LLC under U.S. DOE Contracts DE-AC05-06OR23177 and
DE-AC02-06CH11357. The U.S. Government retains a non-exclusive, paid-up, irrevocable, world-wide license
to publish or reproduce this manuscript for U.S. Government purposes.

%\vspace* {-.17in}

\end{document}